\documentclass[twocolumn,preprintnumbers,amsmath,amssymb,showpacs]{revtex4-1}
\usepackage{graphicx}
\usepackage{dcolumn}
\usepackage{bm}
\usepackage{lipsum}

\begin{document}

\title{Beyond Kerker's conditions: simultaneously nearly zero forward and nearly zero backward scatterings}
\author{Jeng Yi Lee$^{1,2}$, Andrey E. Miroshnichenko$^{3}$, and Ray-Kuang Lee$^{2,4,5,6}$}

\affiliation{
$^1$ Department of Applied Science, National Taitung University, Taitung 95092, Taiwan\\
$^2$ Institute of Photonics Technologies, National Tsing Hua University, Hsinchu 30013, Taiwan\\
$^3$School of Engineering and Information Technology, University of New South Wales, Canberra, Australian Capital Territory 2600, Australia\\
$^4$Physics Division, National Center for Theoretical Sciences, Hsinchu 30013, Taiwan\\
$^5$ Department of Physics, National Tsing Hua University, Hsinchu 30013, Taiwan\\
$^6$Corresponding author: rklee@ee.nthu.edu.tw
}
\date{\today}

\begin{abstract}
With theoretical analyses and numerical calculations, we show that a passive scatterer at the sub-wavelength scale can simultaneously exhibit both nearly zero forward scattering (NZFS) and nearly zero backward scattering (NZBF).
It is related to the interference of dipolar quadrupole modes of different origin, leading to coexistence of Kerker's first and second conditions at the same time.
For optical frequencies, we propose two different sets of composited materials in multi-layered nano-structures, i.e., $CdTe/Si/TiO_2$ and $TiO_2/Au/Si$, for the experimental realization.
\end{abstract}

\maketitle
Scattering fields generated from a localized and finite-sized passive object can be well described by the multipole decomposition~\cite{jackson,multipole}. 
A deep understanding on the  constitutions of  these multipole moments provides the ability to tailor and control the scattered field.
In particular, through the interference between electric and magnetic dipoles, Kerker \textit{et al.}~~\cite{kerker} revealed the possibility to have an asymmetric field radiation with zero backward scattering (ZBS) or zero forward scattering (ZFS), which are known as the first and second Kerker's conditions, respectively.
Even though the power conservation and optical theorem limit the realization of perfect Kerker's conditions~\cite{opticaltheorem1,opticaltheorem2,kerker1}, these directional scattering patterns have been applied to unidirectional emission control ~\cite{antenna1}, as well as  metasurfaces~\cite{brewster,huygen} and metadevices ~\cite{antenna2,antenna3}, with recent experimental observations in nano-particles with high refractive index~\cite{nanoparticle1,nanoparticle2,nanoparticle3} and magnetic spherical particles~\cite{magnetic1}.

Nevertheless, both Kerker's conditions are realized separately. 
In this Letter, we show that with higher order modes (i.e. quadrupoles), one can achieve these two conditions simultaneously, resulting in a passive scatterer exhibiting both NZFS and NZBS.
To go beyond Kerker's conditions, we need to have superposition of a dipole and  quadrupole of different origin with a ratio  $3/5$ in the modulus, as well as a phase difference $\pi$. 
This result can be envisioned from the parity symmetry of corresponding modes, as shown in Figs. 1 (a)-(b).
We will discuss them in detail later.
Through the phase diagram~\cite{kerker1,phase1}, two different sets of composited materials in a three-layered nano-sphere, i.e., $CdTe/Si/TiO_2$ and $TiO_2/Au/Si$ (from outer to inner regions),  are found for the implementation in optical frequencies. 
The results revealed here can be readily served as a building block for directional emission controller, nano-antenna and metasurfaces.

Without loss of generality, we consider a spherical object, i.e., the scatterer, under an electromagnetic plane wave illumination, which has the form $e^{i\,k_{0}\,z-i\, \omega\,t}$  with the polarization of electric field in $\hat{x}$-direction.
Here $k_0$ is the environmental wavenumber and $\omega$ denotes the angular frequency of this plane wave.
Through spherical multipole decomposition, all fields can be constructed by proper choices of wave basis and coefficients \cite{book,phase1}.
Under the excitation of an incident field, power assignments in a passive system would be allocated for the radiative loss (scattering cross section, $\sigma^{scat}$) and the material dissipation loss (absorption cross section, $\sigma^{abs}$), along with the sum of them being the  extinction cross section, i.e., $\sigma^{ext}=\sigma^{scat}+\sigma^{abs}$.
Even unexpected scattering phenomena can be found by embedding specific electromagnetic materials,  power conservation law still exists intrinsically.
In the representation of phasor for the scattering coefficient, we have  $\sigma^{abs}_{n}\geq 0$,  with $\sigma^{abs}_{n}$ denoting the partial absorption cross section.
Consequence of this inequality results in a phase diagram shown in Fig. 1 (c), revealing all unexpected and interesting scattering results~\cite{phase1}.

On the other hand, extinction cross section is also related to the forward scattering amplitude $\vec{f}(\theta=0)$, i.e., $\sigma^{ext}=4\pi/k_{0}\text{Im}[\hat{x}\cdot\vec{f}(\theta=0)]$.
This is a manifestation of the optical theorem \cite{optical,kerker1}.
The scattering field in far-field regime then can be approximated as $E_{0}\vec{f}(\theta,\phi)e^{ik_0 r}/r$, with $E_0$ being the  incident wave amplitude.
It means that the radiative loss $\sigma^{scat}$ and material dissipation loss $\sigma^{abs}$ both contribute to the forward scattering amplitude.
Due to positivity of $\sigma^{abs}$ and $\sigma^{scat}$, thus, a zero value in the forward scattering amplitude imposes zeros both in the absorption and scattering cross sections. 
Correspondingly, forming zero forward scattering is impossible for passive systems.
In addition, within a constant value of extinction cross section, the passive system can display  same scattering amplitude in the forward direction, but with totally different absorption cross section involved \cite{kerker1}.

\begin{figure*}[t]
\includegraphics[width=18.0cm]{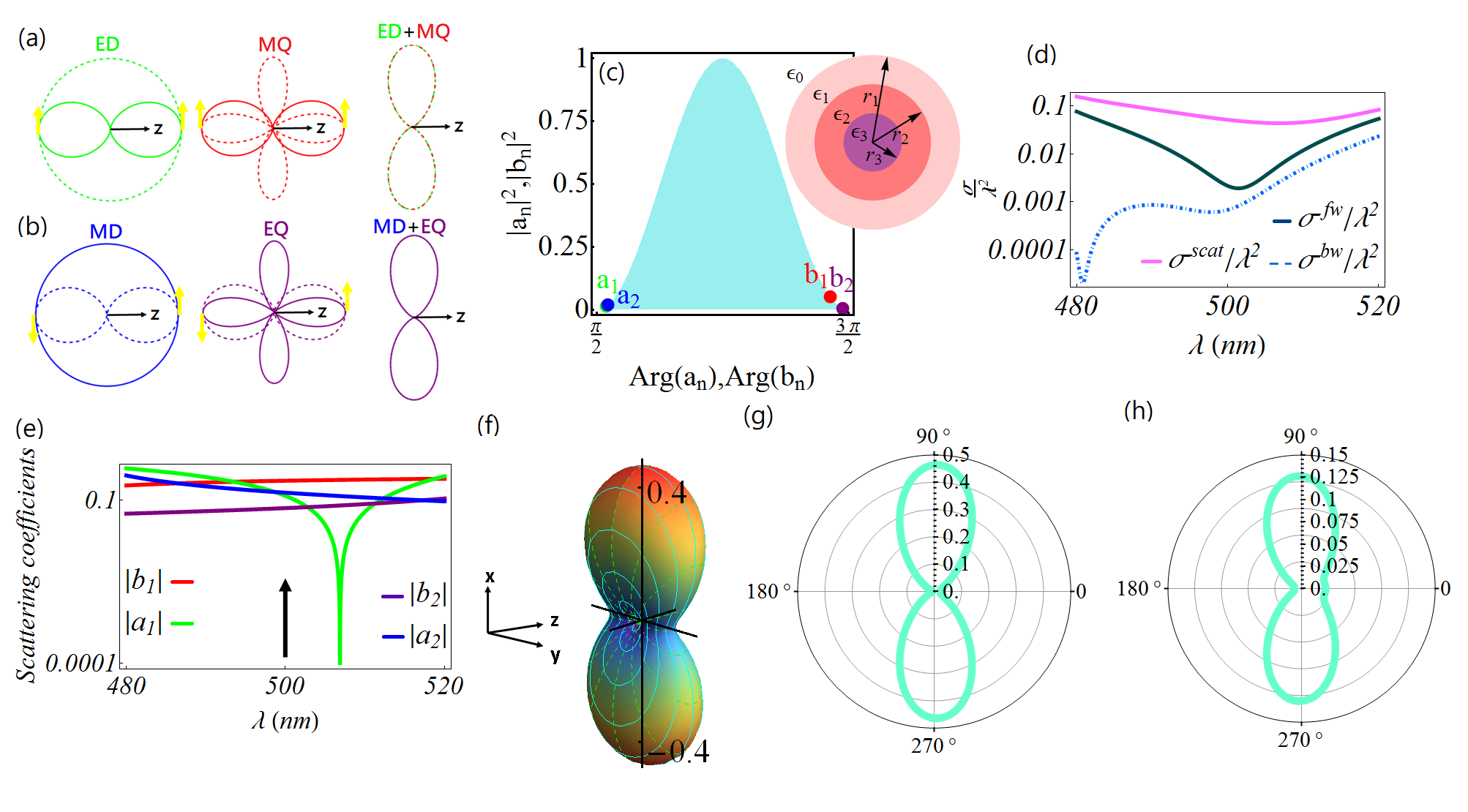}
\caption{(a)-(b) Scattered electric field pattern for electric dipole (ED), magnetic dipole (MD), electric quadrupole (EQ), and magnetic quadrupole (MQ)  in $\hat{x}$-$\hat{z}$ plane (solid line) and in $\hat{y}$-$\hat{z}$ plane (dash line). Yellow arrows denote the scattered electric fields in forward and backward directions, whose upper and downward represent out of phase and in phase to incident electric field. (c) Localizations  in phase diagram for the dominant four multipoles: $a_1$, $b_1$, $a_2$, and $b_2$. Inset depicts our studying system, which is  a three-layered nano-sphere. (d) Forward, backward, and total scattering cross sections, as a function of incident wavelength from $480$ to $520$nm. (e) The corresponding spectra for the absolute values of these four lowest multipole moments. 
At the operation wavelength $\lambda=500$nm, the resulting (f)-(h)  3D  and 2D radiation patterns at (g) $\varphi=0$ and (h) $\varphi=\pi/2$, along $\hat{x}$-$\hat{z}$ plane and $\hat{y}$-$\hat{z}$ plane, respectively.}
\end{figure*}

To tailor the far-field scattering distribution, the corresponding differential scattering cross section can be written as~\cite{book,opticaltheorem1,non-rayleigh}:
\begin{equation}
\frac{d\sigma^{scat}}{d\Omega}=\frac{1}{k_0^{2}}\{\cos^{2}\phi\vert S_{\parallel}(\theta) \vert^2+\sin^{2}\phi\vert S_{\perp}(\theta) \vert^2\}.
\end{equation}
Here, $S_{\parallel}(\theta)$ and $S_{\perp}(\theta)$ are polarized scattering waves parallel and perpendicular to the scattering plane, respectively. Explicitly, the two polarized scattering waves have the form:
\begin{equation}
\begin{split}
S_{\parallel}(\theta)&\equiv\sum_{n=1}^{n=\infty}\frac{2n+1}{n(n+1)}[a_{n}\frac{dP_{n}^{(1)}(\cos\theta)}{d\theta}+b_{n}\frac{P^{(1)}_{n}(\cos\theta)}{\sin\theta}],\\
S_{\perp}(\theta)&\equiv\sum_{n=1}^{n=\infty}\frac{2n+1}{n(n+1)}[b_{n}\frac{dP_{n}^{(1)}(\cos\theta)}{d\theta}+a_{n}\frac{P^{(1)}_{n}(\cos\theta)}{\sin\theta}],\\
\end{split}
\end{equation}
with the associated Legendre polynomial  $P_{n}^{(1)}(\cos\theta)$. The complex scattering coefficients for TM and TE modes are denoted as $a_{n}$ and $b_{n}$, respectively.

To go beyond conventional Kerker's conditions, we are looking for zero scattering in a desired direction $\theta$ by requiring $S_{\parallel}(\theta)=0$ and $S_{\perp}(\theta)=0$.
Owing to the rotation symmetry in our scattering system, along with a  plane wave excitation, here we only have a dependence on the azimuthal angle $\theta$.  
Assume the  dominant excitations are the lowest four multipole moments, i.e., $a_1$ (electric dipole), $b_1$ (magnetic dipole), $a_2$ (electric quadrupole), and $b_2$ (magnetic quadrupole). 
Then, to  eliminate the scattering wave in a desired direction $\theta$, the required conditions become:
 \begin{eqnarray}
3b_1+3a_1\cos\theta=-5b_2\cos\theta-5a_2(1-2\sin^{2}\theta),\\
3b_1\cos\theta+3a_1=-5b_2(1-2\sin^{2}\theta)-5a_2\cos\theta.
 \end{eqnarray}

If only dipole terms are involved, $a_{2} = b_{2} = 0$, to support $S_{\parallel}(0) = S_{\perp}(0)=0$ denoted as ZFS or $S_{\parallel}(\pi) = S_{\perp}(\pi)=0$ denoted as ZBS, we find $a_1=-b_1$ or $a_1=b_1$, respectively.
For a homogeneous sphere,  the conditions $a_1 = -b_1$ or $a_1 = b_1$, lead to the famous formula  $\epsilon=(4-\mu)/(2\mu+1)$ or $\mu=\epsilon$ for ZFS or ZBS, respectively~\cite{kerker}.
Nevertheless, as indicated in phase diagram, see Fig. 1(c), it is impossible to meet a out-of-phase ($\pi$ phase) condition due to the intrinsic constrain on the power conservation.
Moreover, it is obvious that  only with electric and magnetic dipoles,  one  can not satisfy Eqs.  (3) and (4) at the same time.

\begin{figure*}[t]
\includegraphics[width=18.0cm]{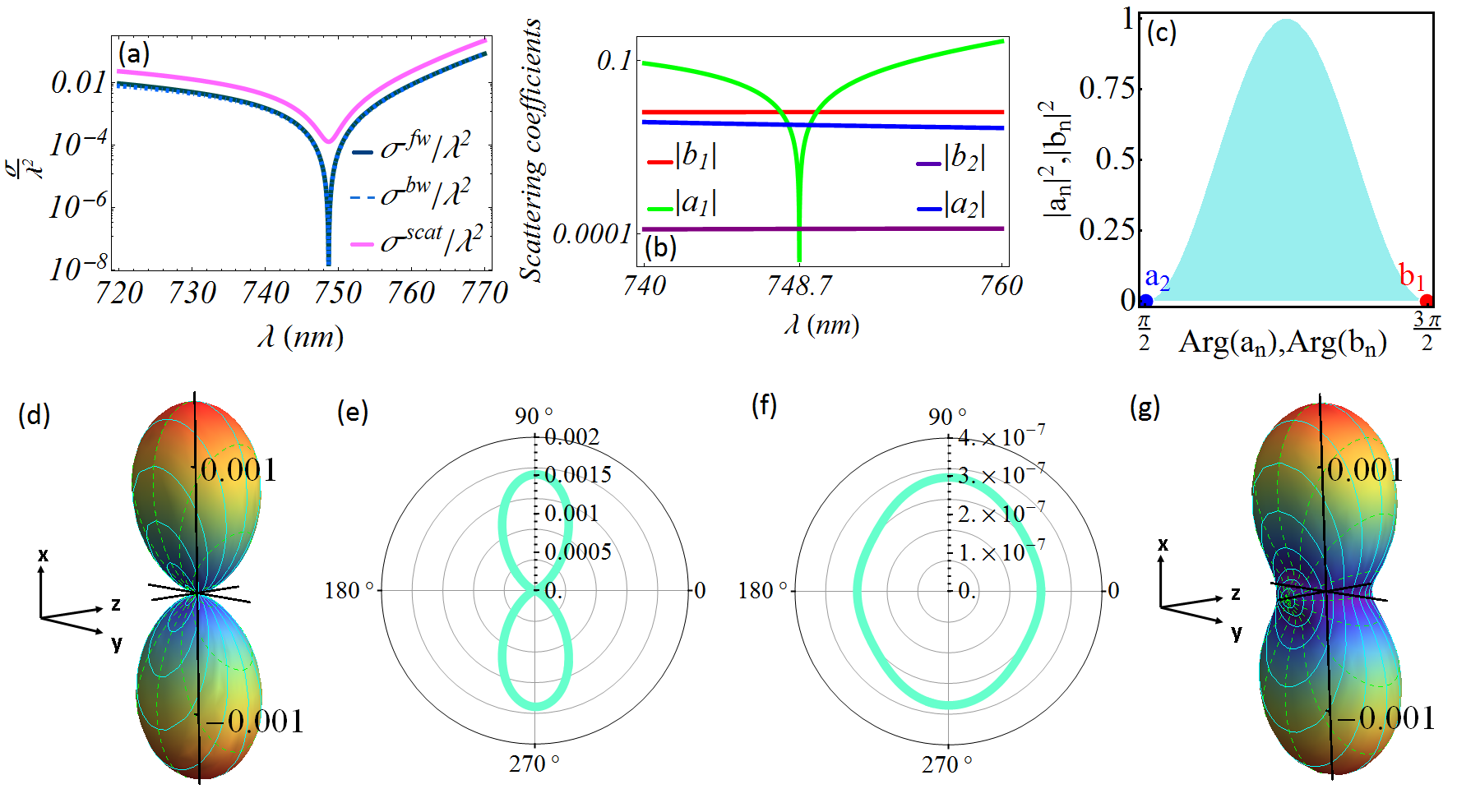}
\caption{(a) Spectra for forward, backward, and total scattering cross sections within $\lambda=720$ to $770$nm; and  (b) spectra for the absolute values of four dominant scattering coefficients, with (c) their  locations in phase digram. At the operation wavelength $\lambda=748.7$nm, the resulting (d) 3D and 2D radiation patterns,  along (e) $\hat{x}$-$\hat{z}$ plane and  (f) $\hat{y}$-$\hat{z}$ plane. If material loss is taken into account, the corresponding 3D radiation pattern is depicted in (g).}
\end{figure*}

Now, by considering the electric and magnetic quadrupoles, from Eqs. (3-4) the conditions to simultaneously suppress scattering field in the forward ($\theta=0$) and backward ($\theta=\pi$) directions can be found as:
\begin{eqnarray}\label{condition}
a_{1}=-\frac{5}{3}b_{2}\,; \qquad b_{1}=-\frac{5}{3}a_{2}.
\end{eqnarray}
This simple result reveals that even for the complicated interferences among four multipole moments, to achieve ZFS and ZBS simultaneously, a required phase difference $\pi$ is asked for the corresponding  electric dipole (magnetic dipole) to the magnetic quadrupole (electric quadrupole), along with a ratio of modulus by $5/3$.

To indicate physics behind our finding in Eq.(\ref{condition}), we employ the phase analysis of multipole moments with highlighting the parity in forward and backward directions \cite{liu}. 
For the first (second) case of Eq.(\ref{condition}), i.e., with electric dipole (magnetic dipole) and magnetic quadrupole (electric quadrupole), in Figs. 1(a)-(b), one can observe their phase parity in forward and backward  directions, which can enable to form wave destructive to achieve ZFS and ZBS simultaneously.
Note that we find the destructive cancellation of resultant scattered  field in $\hat{x}$-$\hat{z}$ plane ($\hat{y}$-$\hat{z}$ plane), but that in $\hat{y}$-$\hat{z}$ plane ($\hat{x}$-$\hat{z}$ plane) possesses ZFS and ZBS.

On the other hand, from the phase diagram, we know that one can only approach the phase difference $\pi$ asymptotically, resulting in nearly zero forward scattering (NZFS) and nearly zero backward scattering (NZBF). By substituting Eq. (5) into Eq. (1), the corresponding scattering cross-session has the form: 
\begin{equation}
\frac{d\sigma^{scat}}{d\Omega}\sim\cos^{2}\phi\sin^{4}\theta\vert a_2\vert^{2}+\sin^{2}\phi\sin^{4}\theta\vert b_2\vert^{2},
\end{equation}
which has zero scattering both in the backward ($\theta=\pi$) and forward ($\theta=0$) directions.
We remark that the role of $a_2$ ($b_2$) is to assign the scattering field along  $\hat{x}$ ($\hat{y}$)-direction for the prefactor $\cos^{2}\phi$ ($\sin^{2}\phi$).

To realize NZFS and NZBS simultaneously, we take a three-layered nano-sphere as the possible experimental platform, which can be composited by  cadmium telluride ($CdTe$) in shell,  silicon ($Si$) in middle, and titanium dioxide ($TiO_2$) in core.
It is noted that these materials have  high refractive index (HRI), in order to induce quadrupole moments and magnetic response at the sub-wavelength scale.
Extra benefit from HRI materials includes a negligible material dissipation loss~\cite{nanoparticle3}.
As illustrated in the inset of Fig. 1(c), the corresponding radius are chosen as $r_1=100$nm, $r_2=87.7$nm, and $r_3=26.3$nm.
With current state of the art technologies, this three-layer nano-sphere can be synthesized by chemical reaction method with a  self-sacrificing template~\cite{chemical1,chemical2}.

By taking the dispersion relations of these three materials into account~\cite{handbook}, we report the log-plot for forward, backward, and total scattering cross sections in the optical wavelengths, $\lambda=480$ to $520$nm, in Fig. 1(d).
As one can see that both the  forward and backward scattering cross sections are much smaller than that of the total scattering cross section.
The corresponding  absolute values for the lowest 4 multipole moments, $a_1$, $a_{2}$, $b_{1}$, and $b_{2}$, are also depicted in Fig. 1(e) in green, blue, red, and purple curves, respectively.
In particular, at $\lambda=500$nm, the corresponding multiple moments, with their locations  marked in Fig. 1(c), can meet the conditions given in  Eq.(5) approximately.
The resulting 3D and 2D radiation patterns are shown in Figs. 1(f) and (g-h), respectively, as a clear demonstration for supporting NZFS and NZBS simultaneously. 
Here the parameters we use are $\epsilon_1=9\epsilon_0$, $\epsilon_2=18.46\epsilon_0$, and $\epsilon_3=6.5\epsilon_0$.

In addition to have 4 lowest multiple moments, another possibility to have NZFS and NZBS simultaneously is to have electric quadrupole $a_2$ and magnetic dipole $b_1$ only, with $a_1 = b_2 = 0$.
To excite the magnetic dipole, we use titanium dioxide ($TiO_2$) in shell, gold ($Au$) in middle, and silicon ($Si$) in core for our three-layered nano-sphere.
Now, the geometrical size parameters are chosen as $r_1=85$nm, $r_2=72.3$nm, and $r_3=43.4$nm.
Even though gold is a lossy material, at the first stage we embed material dispersions of these  three composites into our calculation but without lossy terms~\cite{handbook}.

In Fig. 2(a), we report the corresponding spectra for forward, backward, and total scattering cross sections from our proposed structure.
As one can see clearly, there is a dip at $\lambda=748.7$nm, at which both forward and backward scattering cross section approach zeros.
The corresponding spectra for the lowest four multipole moments are also depicted in Fig. 2 (b).
As we design, the electric dipole  ($a_1$) is minimum at $\lambda=748.7$nm; while the  dominant channels become $b_1$ and $a_2$.
The locations of $b_1$ and $a_2$ are shown in Fig. 2(c), which indicates a nearly $\pi$ phase  difference between the magnetic dipole and electric quadrupole.

The resulting radiation patterns are illustrated in 3D plot, Fig. 2(d) and in 2D plots with $\varphi=0$ ($\hat{x}$-$\hat{z}$ plane), Fig. 2(e) or with $\varphi=\pi/2$ ($\hat{y}$-$\hat{z}$ plane), Fig.  2(f).
The electrical permittivities used here are  $\epsilon_1=6.25\epsilon_0$, $\epsilon_2=-20.9\epsilon_0$, and $\epsilon_3=13.95\epsilon_0$ for the shell, middle, and core regions, respectively.
When the material loss is taken into consideration, the required $\pi$ phase difference will be destroyed, resulting in some uncanceled   forward and backward scatterings, as illustrated in Fig. 2(g).
We observe that with the lossy effect, the radiation pattern still remains a similar shape as that of a lossless one,  but significant scatterings can be found in all directions.

Before conclusion, we want to remark that achieving such anomalous scatterers with NZFS and NZBS simultaneously, the material system can be composited without plasmonics.
However, embedding HRI material in subwavelengthly scatterers is necessary, in order to excite quadrupole moments and magnetic response.
As we known from optical theorem, it is impossible to meet the required conditions given in Eq. (5) perfectly. However,  approaching  a $\pi$ phase difference between the scattering coefficients still gives useful consequences.
Even though, there are always other-order terms contributing to the end result, our direct numerical calculations illustrate good agreement with analytical results only with few lowest orders. Note also, that unwanted dipole response, for example, can be completely suppressed due to anapole type states excitations~\cite{Miroshnichenko:NatComm:2015} 

In conclusion,  we provide the required conditions to eliminate scattering at an arbitrary direction with the interferences among electrical/magnetic dipoles and quadrupoles.
To go beyond Kerker's conditions, we reveal the required conditions to  support nearly zero forward scattering (NZFS) and nearly zero backward scattering (NZBF) simultaneously.
By considering a three-layered nano-sphere at the sub-wavelength scale, we propose two different sets of composited materials, i.e., $CdTe/Si/TiO_2$ and $TiO_2/Au/Si$ as the shell/middle/core, for the experimental implementation in optical wavelengths.
Our results  can be readily applied  for nano-antenna, directional emission control, and metasurfaces.

\section*{Funding.} Ministry of Science and Technology, Taiwan (MOST) (105-2628-M-007-003-MY4).

\bibliography{achemso-demo}

\end{document}